\documentclass[aps,twocolumn,showpacs,preprintnumbers,amsmath,amssymb,floatfix]{revtex4}
\usepackage{graphicx}
\usepackage{dcolumn}
\begin{document}
\title{On-line trading as a renewal process: Waiting time and inspection paradox}
\author{Jun-ichi Inoue}
\email{j_inoue@complex.eng.hokudai.ac.jp}
\affiliation{
Complex Systems Engineering, Graduate School of Information
Science and Technology, Hokkaido University,
N14-W9, Kita-ku, Sapporo 060-0814, Japan}
\author{Naoya Sazuka}
\email{Naoya.Sazuka@jp.sony.com}
\affiliation{
System Technologies Laboratories, 
Sony Corporation, 5-1-12 Kitashinagawa Shinagawa-ku, 
Tokyo 141-0001, Japan}
\author{Enrico Scalas}
\email{enrico.scalas@mfn.unipmn.it}
\affiliation{
Dipartimento di Scienze e Technologie, 
Universit\`a del Piemonte Orientale, Viale T. Michel, 11 I-15121
Alessandria, Italy}
\date{\today}
\begin{abstract}
We briefly review our recent studies on 
stochastic processes modelling internet on-line trading.  
We present a way to evaluate the average waiting time
between the observation of the price in financial markets
and the next price change, 
especially in an on-line foreign exchange trading service
for individual customers
via the internet.
The basic method  
of our approach depends on 
the so-called renewal-reward theorem. 
Assuming that the stochastic 
process modelling the price change
is a renewal process, 
we use the theorem to calculate the average waiting time 
of the process. 
The so-called ``inspection paradox'' is discussed, which, in general, means that the average
durations is shorter than the average 
waiting time. 
\end{abstract}
\maketitle
\section{Introduction}
\label{sec:Intro}
Financial data have attracted a lot 
of attention from physicists as informative material to 
investigate the macroscopic behavior of markets \cite{Mantegna2000,Bouchaud,Voit}. 
Some of these studies are restricted to 
the stochastic variables of the price changes 
(returns) and most of them concern a key-word: {\it Fat tails} of the 
distributions \cite{Mantegna2000}. However, also 
the distribution of time intervals 
can deliver useful information on the 
markets and it is worth while to investigate these 
properties extensively 
\cite{Engle98,Mainardi,Raberto,Scalas,Kaizoji,Scalas06} 
and if possible, to apply the gained knowledge to financial engineering. 

Fluctuations 
in time intervals between events 
are not only peculiar to financial markets, but also very common in science.
For instance, the spike train of a single neuron is characterized
by a time series in which the time difference between consecutive
spikes is not constant but fluctuates. 
This stochastic process, specified by the so-called 
Inter-Spike Intervals (ISI), 
is one of such examples \cite{Tuckwell,Gerstner}. 
The average of 
the ISI is about a few milli-seconds 
and the distribution of the durations (intervals) is 
well-described by the {\it Gamma distribution} \cite{Gerstner}. 

On the other hand, 
in financial markets, 
for instance, the time intervals of 
two consecutive transactions 
of BUND futures (BUND is the German word for bond)
and BTP futures
(BTP is the middle and long term Italian Government bonds 
with fixed interest rates)
traded at LIFFE 
(LIFFE stands for London International Financial Futures and Options Exchange) 
are $\sim 10$ seconds 
and are well-fitted by the so-called {\it Mittag-Leffler 
distribution} \cite{Mainardi,Raberto,Scalas}. 
The Mittag-Leffler 
distribution behaves as a stretched exponential 
distribution in the 
short interval regime, 
whereas for the long interval regime, 
the function has a power-law tail. 
Therefore, the behavior of the distribution 
described by the Mittag-Leffler function 
varies from stretched 
exponential to power-law at some intermediate critical interval 
\cite{Gorenflo}. 

The Sony bank USD/JPY exchange rate \cite{Sony},
(i.e. the rate
for individual customers
of the Sony bank
in their on-line foreign exchange trading service) 
is a good example to be checked against 
the Mittag-Leffler distribution.
Actually, our results implied that 
the Mittag-Leffler distribution 
does not well fit the Sony bank rate \cite{SazukaInoue2006}. 
The Sony bank rate has 
$\sim 20$ minutes \cite{Sazuka} 
as the average time interval 
which is much longer than other 
market time intervals such as those for the BUND future. 
This is due to the fact that the the Sony back rate 
can be regarded as a so-called {\it first-passage process} 
\cite{Redner,Kappen,Gardiner,Risken,Simonsen,Kurihara}
for raw market data. 
\begin{table}
\caption{\label{tab:table0}
Three typical examples with fluctuation between 
events.
}
\begin{tabular}{|c||c|c|c|}
\hline 
\mbox{} & 
ISI & BUND future & Sony bank rate  \\
\hline
Average duration 
& $\sim 3$ [ms]  & $\sim 10$ [s] & $\sim 20$ [min] \\
\hline
PDF & Gamma & 
Mittag-Leffler & Weibull \\
\hline
\end{tabular}
\end{table}
In Table \ref{tab:table0}, we list 
the average time intervals and 
the probability distribution function 
(PDF) that describes the data 
with fluctuation between the events for 
typical three examples, 
namely, 
the ISI, the BUND future and 
the Sony bank rate. 
From this table, 
an important question 
arises. 
Namely, how long do the customers 
of the Sony bank should wait
between observing the price 
and the next price change? 
This type of 
question never occurs
in the case of the ISI or of
the BUND future, because 
the average time intervals are too short to 
evaluate such informative measure. 

For the customers, an important 
(relevant) quantity is the waiting time rather than the time interval  
between consecutive rate changes. Here, the waiting time
is defined as the time the customer has to wait between the instant in which they 
enter the market in the World-Wide Web and the 
next price change \cite{Sony}. 
If the sequence of time intervals is non-exponential and 
the customers observe the rate at random on the time axis, 
the distribution of the waiting time no longer coincides 
to the distribution of the time intervals. 

In this review article, 
we present a useful way to evaluate moments of arbitrary order 
for the waiting time 
and for  arbitrary duration distribution of price changes 
as well as observation time distribution, by 
directly deriving the waiting time distribution. 

This paper is organized as follows. In the next section \ref{sec:SBR}, 
we introduce the Sony bank rate \cite{Sony} 
which is generated from 
the high-frequency foreign exchange market rate
via the rate window with width $2\epsilon$ yen 
($\epsilon=0.1$ yen for the Sony bank). 
Then in the subsequent section \ref{sec:Deriv}, we explain our method 
to derive waiting-time distributions. We show that our treatment 
reproduces the result obtained by the renewal-reward theorem. 
We also evaluate the deviation around the average waiting time for 
the Weibull first-passage time distribution and uniform observation 
time distribution. We find that the resultant standard deviation is 
the same order as the average waiting time. 
We test our analysis for several observation-time 
distributions and calculate higher-order moments. 
The so-called ``inspection paradox'' \cite{Angus} is mentioned in section IV, which means in general that the average of 
durations is shorter than the average 
waiting time. 
The last section \ref{sec:Summary} contains concluding remarks. 
\section{The Sony bank rate: An example of first-passage process}
\label{sec:SBR}
\begin{figure}[ht]
\begin{center}
\includegraphics[width=8.5cm]{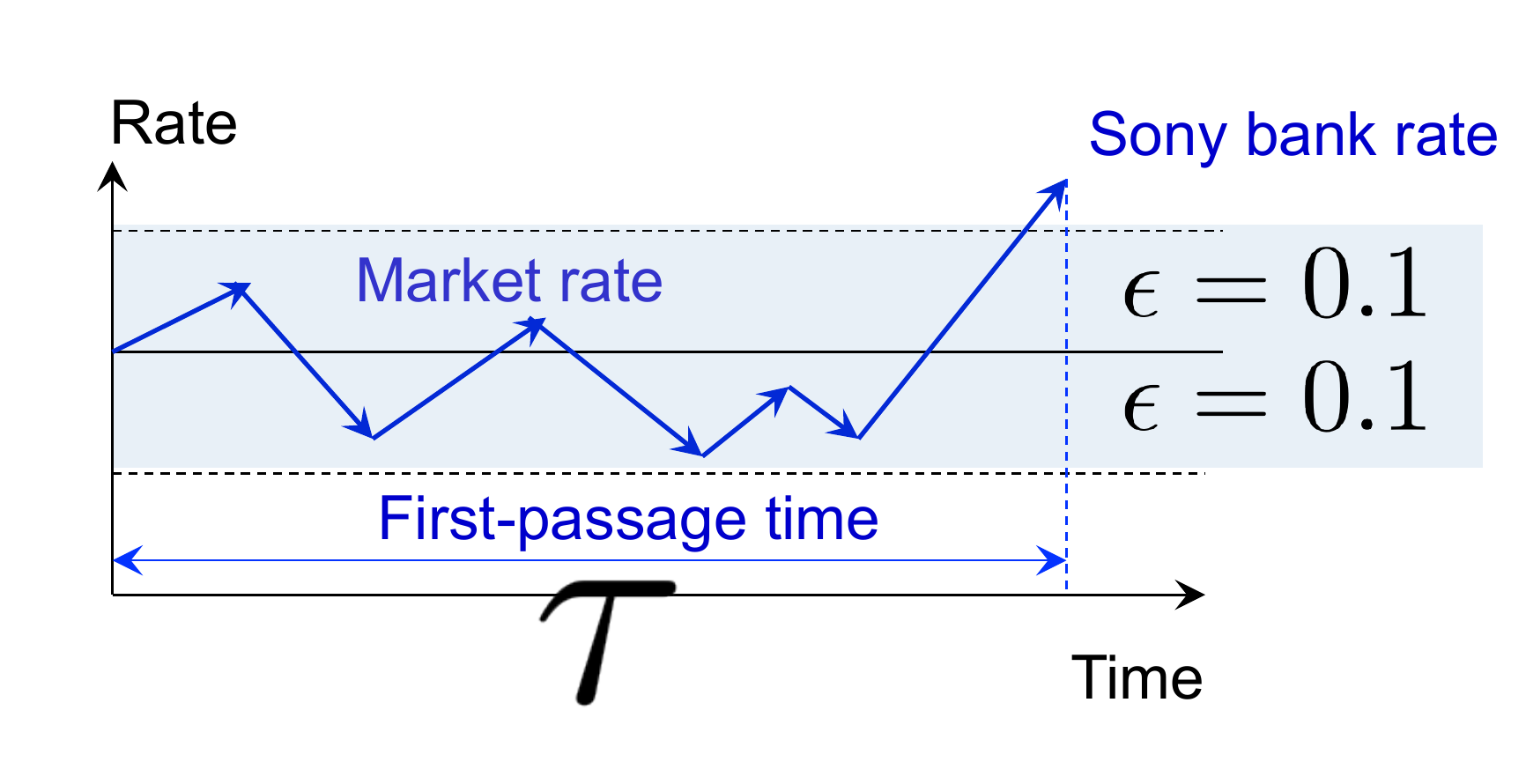}
\end{center}
\caption{\footnotesize An illustration of 
generating the filtered rate 
by the 
rate window with width $2\epsilon$ from the 
market rate. 
If the market rate changes by a quantity 
greater or equal to $0.1$ yen, 
the Sony bank USD/JPY exchange 
rate is updated to the market rate. 
}
\label{fig:fg_window}
\end{figure}

The Sony bank rate 
we deal with in this paper 
is the rate
for individual customers
of the Sony bank \cite{Sony} 
in their on-line foreign exchange trading service
via the internet. 
If the USD/JPY market rate changes by 
greater or equal to $0.1$ yen, 
the Sony bank USD/JPY exchange 
rate is updated to the market rate. 
In this sense, the Sony bank rate 
can be regarded as a first 
passage processes \cite{Redner,Kappen,Gardiner,Risken,Simonsen,Kurihara}. 
In Fig. \ref{fig:fg_window}, 
we show the mechanism of generating the Sony bank rate 
from the market rate (this process is sometimes 
referred to as a
{\it first exit process} \cite{Montero}). 
As shown in the figure, 
the time difference between 
two consecutive points in the Sony bank rate 
becomes longer than the time intervals of the market rates. 
In Table \ref{tab:table1}, 
we show 
several data concerning 
the Sony bank USD/JPY 
rate vs. tick-by-tick 
data by Bloomberg for USD/JPY rate. 
\begin{table}
\caption{\label{tab:table1}
The Sony bank USD/JPY exchange rate vs. tick-by-tick 
data for USD/JPY exchange rate. }
\begin{tabular}{|c|c|c|}
\hline
\mbox{} & Sony bank rate & tick-by-tick data \\
\hline
$\#$ of data a day & $\sim 70$  & $\sim 10,000$ \\
\hline
The smallest price change  & $0.1$ yen  & $0.01$ yen \\
\hline
Average duration & $\sim 20$ minutes & $\sim 7$ seconds \\
\hline
\end{tabular}
\end{table}
It is a non-trivial problem to ask what kind of 
distribution is suitable to explain the distribution 
of the first-passage time. 
For this problem, 
we attempted to check several statistics 
from both the analytical and the empirical points of view  
under the assumption that 
the first-passage time might obey 
a Weibull 
distribution \cite{Sazuka0,Sazuka2,InoueSazuka2006}. 
We found that 
the data are well fitted by 
a Weibull distribution. 
This fact means that 
the difference between successive rate changes in Sony bank 
fluctuates and has some memory.  
\section{Derivation of the waiting time distribution}
\label{sec:Deriv}
In this section, we derive the distribution 
of the waiting time for the customers. 
Our approach enables us to 
evaluate not only the first moment of the 
waiting time but also moments of any order.  
\subsection{The probability distribution of the waiting time}
\begin{figure}[ht]
\begin{center}
\includegraphics[width=8.5cm]{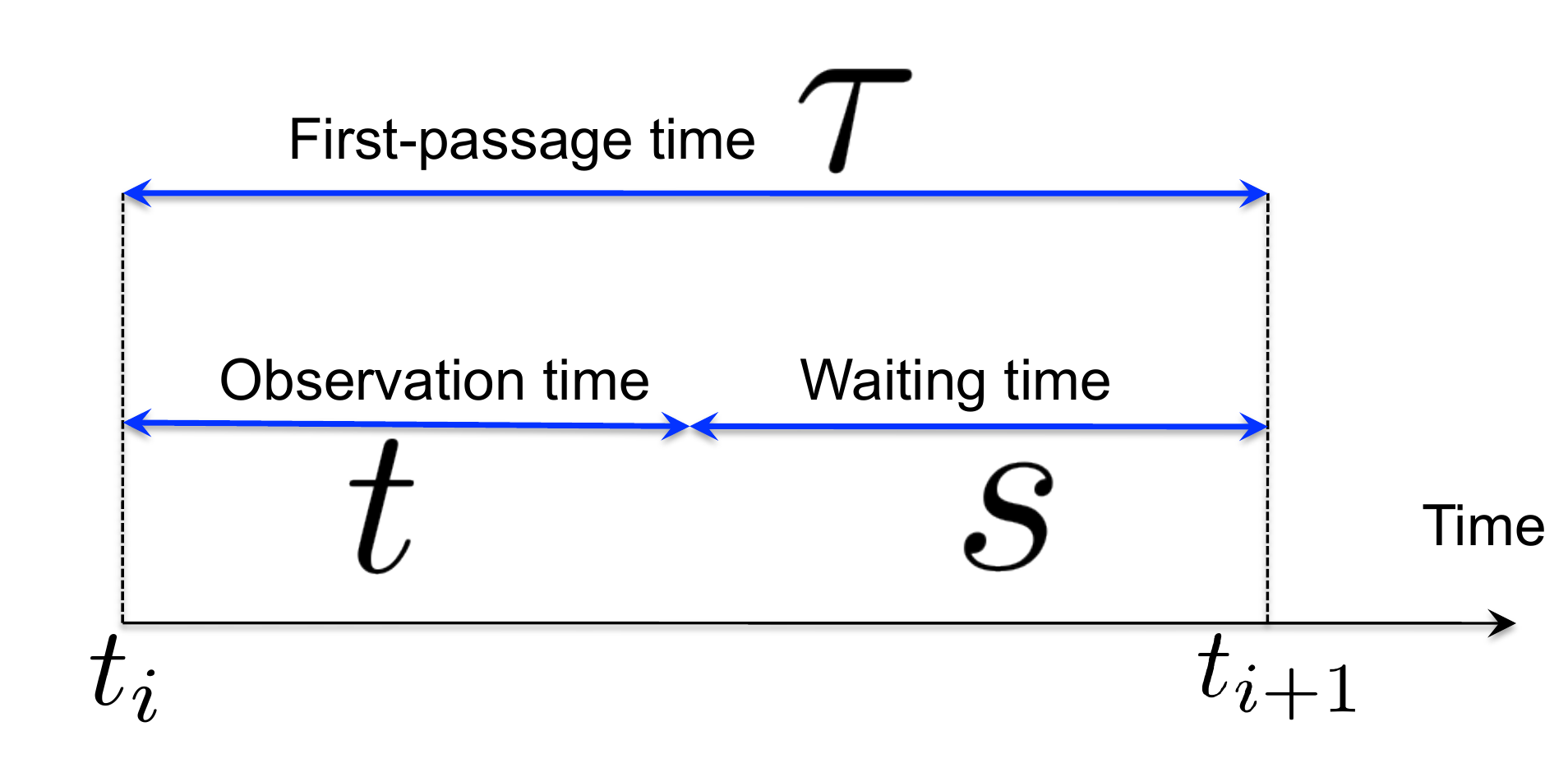}
\end{center}
\caption{\footnotesize 
The relation 
these points 
$\tau,t$ and $s$
in time axis. 
The first-passage time $\tau$ 
is given by 
$\tau=t_{i+1}-t_{i}$. 
The observation time is measured from 
the point $t _{i}$. 
}
\label{fig:fg9b}
\end{figure}

\begin{figure}[ht]
\begin{center}
\includegraphics[width=8.5cm,bb=0 0 360 252]{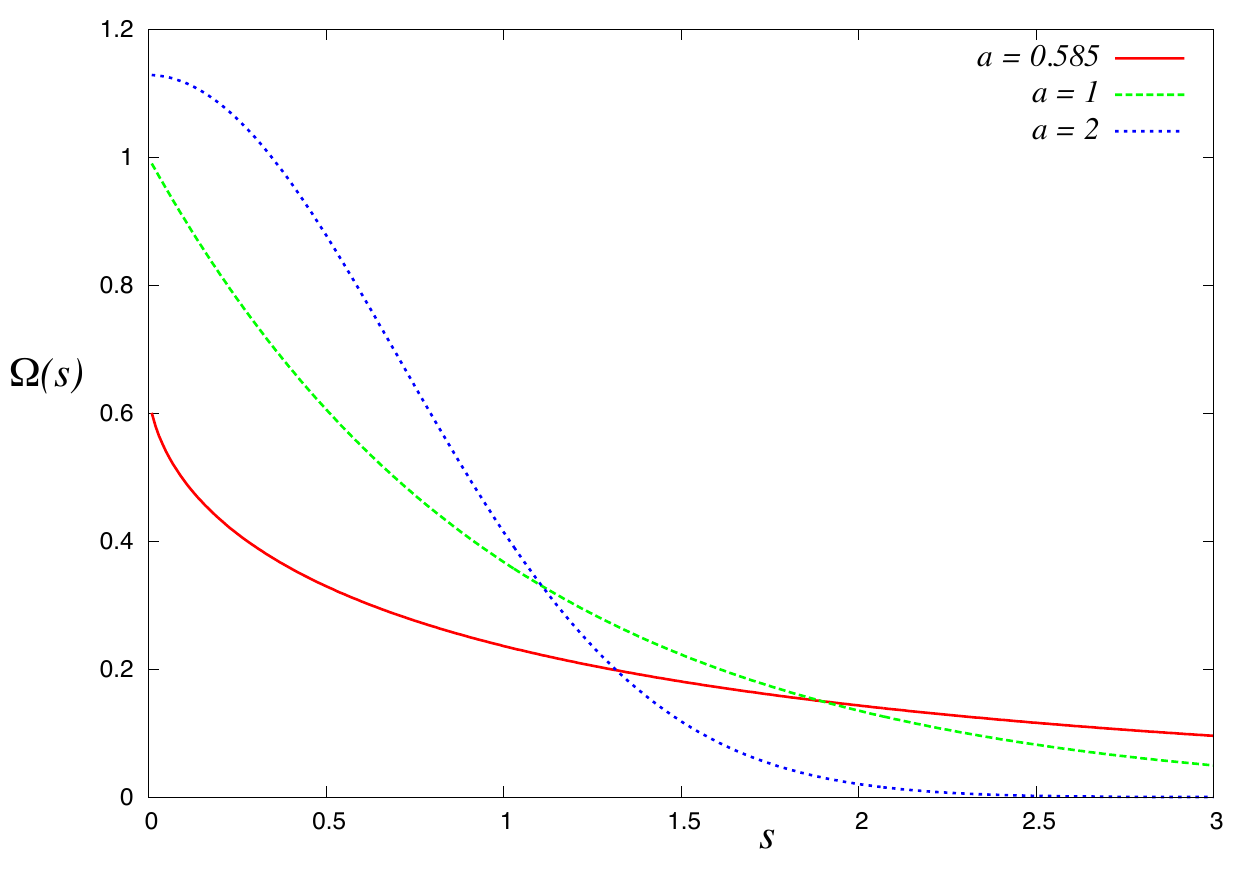}
\end{center}
\caption{\footnotesize 
The distribution 
of waiting time 
for a Weibull distributing 
$\Omega (s)$ with $a=1$ and 
$m=0.59,1$ and $2$. }
\label{fig:fg9c}
\end{figure}

We first derive the probability distribution function of 
the waiting time $s$. 
Let us suppose that 
the difference between 
two consecutive points 
of the Sony bank rate change, 
namely, the first-passage time 
$\tau$ 
follows the distribution with probability density function $P_{W}(\tau)$. 
Then, the customers observe the 
rate at time $t$ 
($0 \leq t \leq \tau$) 
that should be 
measured from the point at which 
the rate previously changed. 
In Fig. \ref{fig:fg9b}, 
we show the relation 
among these variables 
$\tau$, $t$ and $s$
in the time axis. 
The waiting time for 
the customers 
is naturally defined as 
$s \equiv \tau - t$. 
Now, notice that 
the distribution $\Omega (s)$ 
can be written in terms of the first-passage time 
distribution (with density $P_{W}(\tau)$) and the observation time distribution 
(with density $P_{O}(t)$) of the customers as a convolution
\begin{eqnarray}
\Omega (s) & \propto & 
\int_{0}^{\infty}
d\tau 
\int_{0}^{\tau}
dt 
\,Q(s|\tau,t)P_{O}(t)P_{W}(\tau). 
\end{eqnarray}
In this equation, the conditional   
probability density
$Q(s|\tau,t)$ that 
the waiting time 
takes the value $s$ provided that 
the observation time and 
the first-passage time 
were given as $t$ and $\tau$, 
respectively, 
is given by 
\begin{eqnarray}
Q(s|\tau,t) & = & 
\delta (s-\tau+t)
\end{eqnarray}
where
$\delta (\cdot)$ is Dirac's delta function. 
Taking into account the 
normalization constant of 
$\Omega (s)$, we have
\begin{equation}
\Omega (s) =
\frac{\int_{0}^{\infty}
d\tau P_{W}(\tau) 
\int_{0}^{\tau}
dt \,\delta (s-\tau +t) P_{O}(t)} 
{
\int_{0}^{\infty}
ds
\int_{0}^{\infty}
d\tau P_{W}(\tau) 
\int_{0}^{\tau} 
dt \,\delta (s-\tau + t)P_{O}(t)} 
\end{equation}
where, again, $t$ denotes the observation time for the customers. 
The result of the renewal-reward theorem : 
$w=\langle s \rangle = E(\tau^{2})/2E(\tau)$ 
(see for example \cite{Oishi}) is 
recovered by inserting a uniformly distributed observation time 
distribution $P_{O}(t)=1$ 
into the above expression. Indeed, we have
\begin{eqnarray}
w & = & 
\langle s \rangle = 
\int_{0}^{\infty}
ds s \Omega (s) = 
\frac{
\int_{0}^{\infty}
ds s 
\int_{s}^{\infty}
d\tau 
P_{W}(\tau)}
{
\int_{0}^{\infty}
ds 
\int_{s}^{\infty}
d\tau 
P_{W}(\tau)} \nonumber \\
\mbox{} & = & 
\frac{
\int_{0}^{\infty}
\frac{d}{ds}\{s^{2}/2\} ds
\int_{s}^{\infty}
d\tau 
P_{W}(\tau)}
{
\int_{0}^{\infty}
\frac{d}{ds}\{s\} ds 
\int_{s}^{\infty}
d\tau 
P_{W}(\tau)} \nonumber \\
\mbox{} & = & 
\frac{
(1/2) \int_{0}^{\infty}
s^{2}P_{W}(s)ds}
{
\int_{0}^{\infty}
sP_{W}(s)ds} = 
\frac{E(\tau^{2})}{2E(\tau)}
\end{eqnarray}
where we defined 
the $n$-th moment of the 
first-passage time 
$E(\tau^{n})$ by 
\begin{eqnarray}
E(\tau^{n}) & = & 
\int_{0}^{\infty}
ds s^{n} 
P_{W}(s).
\end{eqnarray}
More generally, we may choose a non-uniform
$P_{O}(t)$. 
For this general form of 
the observation time distribution, 
the probability distribution of 
the waiting time $s$ is given as follows. 
\begin{eqnarray}
\Omega (s) & = & 
\frac{
\int_{s}^{\infty}
d\tau 
P_{W}(\tau) P_{O} (\tau-s)}
{
\int_{0}^{\infty}ds
\int_{s}^{\infty}
d\tau 
P_{W}(\tau) P_{O}(\tau-s)} \nonumber \\
\mbox{} & = &
\frac{
\int_{s}^{\infty}
d\tau 
P_{W}(\tau) P_{O} (\tau-s)}
{E(t) - \delta_{1}}
\label{eq:Omega_s}
\end{eqnarray}
where we defined $\delta_{n}$ by 
\begin{eqnarray}
\delta_{n} & = & 
\int_{0}^{\infty}
\frac{ds s^{n}}{n}
\int_{s}^{\infty}
P_{W}(\tau) 
\frac{\partial P_{O} (\tau-s)}
{\partial s}.
\end{eqnarray}
By using the same method as in the derivation of the 
distribution 
$\Omega (s)$, 
we easily obtain 
the first two moments of 
the waiting time distribution as
\begin{eqnarray}
\langle s \rangle & = & 
\frac{E(\tau^{2})/2 - 
\delta_{2}}
{E(\tau)-\delta_{1}},\,\,\,
\langle s^{2} \rangle = 
\frac{E(\tau^{3})/3- 
\delta_{3}}
{E(\tau)-\delta_{1}}
\label{eq:two_Moments}
\end{eqnarray}
and the study of the standard deviation leads to 
\begin{equation}
\sigma = 
\sqrt{
\frac{
\{
4E(\tau^{3})E(\tau) - 
3E(\tau^{2})\}
+G_{\delta_{1},\delta_{2},\delta_{3}}  
}
{
12(E(\tau)-\delta_{1})^{2}
}
}
\label{eq:Deviation}
\end{equation}
\begin{eqnarray}
G_{\delta_{1},\delta_{2},\delta_{3}} & = & 
- 
4\delta_{1} E(\tau^{3}) - 
12\delta_{3}E(\tau)  \nonumber \\
\mbox{} & + & 
12\delta_{2}E(\tau^{2}) + 12\delta_{1} \delta_{3} 
-12\delta_{2}^{2}
\end{eqnarray}
where 
we defined 
\begin{eqnarray}
\langle s^{n} \rangle & = & 
\int_{0}^{\infty}
dss^{n}\Omega (s).
\end{eqnarray}
Thus, this probability distribution $\Omega (s)$ enables us 
to evaluate moments of any order for the waiting time. 
We consider the case of 
$P_{O}(\tau)=1$ as an example. 
This case corresponds to 
the result obtained by the renewal-reward theorem 
\cite{InoueSazuka2006}. 
We find that 
$\delta_{n} =0$ holds for 
arbitrary integer $n$. 
Thus, 
the waiting time distribution 
$\Omega (s)$ leads to 
\begin{eqnarray}
\Omega (s) & = & 
\frac{\int_{s}^{\infty}
P_{W}(\tau)}
{E(\tau)}. 
\end{eqnarray}
Then, the average waiting time and 
the deviation around the value lead to 
\begin{eqnarray}
w & = & 
\frac{E(\tau^{2})}{2E(\tau)},\,\,\,\,
\sigma = 
\sqrt{
\frac{4E(\tau^{3})E(\tau) - 
3E(\tau^{2})^{2}}
{12E(\tau)^{2}}
}.
\label{eq:def_dev}
\end{eqnarray}
For a Weibull distribution 
having the parameters $m,a$, 
the above results can be re-written as 
\begin{eqnarray}
\Omega (s) & = & 
\frac{m \, {\rm e}^{-s^{m}/a}}
{a^{1/m} \Gamma 
\left(
\frac{1}{m}
\right)} \\
w & = & 
a^{1/m}
\frac{\Gamma 
\left(
\frac{2}{m}
\right)}
{
\Gamma 
\left(
\frac{1}{m}
\right)
} \\
\sigma & = & 
\frac{a^{1/m} \sqrt{
\Gamma(1/m)\Gamma (3/m) - 
\Gamma (2/m)^{2}}}
{\Gamma (1/m)}
\end{eqnarray}
where we defined the Gamma 
function as
\begin{eqnarray}
\Gamma(x) & = & 
\int_{0}^{\infty}dt\, 
t^{x-1}{\rm e}^{-t}.
\end{eqnarray}
It is important for us 
to notice that 
for an exponential 
distribution 
$m=1$, 
we have 
$w=\sigma=a$ 
by taking into account the fact that 
$\Gamma (n)=(n-1)!$. 
Moreover,
the average waiting time $w$
is identical to the average time interval
$E(\tau)$
since 
$w=E(\tau^{2})/2E(\tau)=E(\tau)$ 
holds 
if and only if 
$m=1$ (The rate change follows a Poisson 
arrival process).
These results are 
obtained by using a different method in 
our previous studies \cite{InoueSazuka2006}. 
In 
Fig. \ref{fig:fg9c}, 
we plot the 
distribution 
$\Omega (s)$ 
for $m=1,2$ and $m=0.585$. 
\subsection{Comparison with empirical data analysis}
It is time to compare the analytical result with 
that of the empirical data analysis. 
For uniform observation 
time distribution 
$P_{O} (t)=1$, 
we obtained 
$\sigma=60.23$ minutes. 
On the other hand, 
from empirical data analysis, we 
evaluate 
the 
quantity (\ref{eq:def_dev}) 
by sampling 
the moment 
as $E(\tau^{n}) = 
(1/N)\sum_{i=1}^{N}
\tau_{i}^{n}$ directly 
from Sony bank rate data \cite{Sony} 
and find $\sigma = 74.35$ minutes. 
There exists a finite gap between 
the theoretical prediction and the result by 
the empirical data analysis, however, 
both results are of the same order of magnitude. 
The gap might become small if we 
take into account the power-law tail 
of the first-passage time distribution. 
In fact, we showed that 
for the average waiting time, 
the power-law tail makes 
the gap between the theoretical prediction and 
empirical observation smaller \cite{SazukaInoue2006}.  
\section{Inspection paradox}
\begin{figure}[ht]
\begin{center}
\includegraphics[width=8.5cm,bb=0 0 360 252]{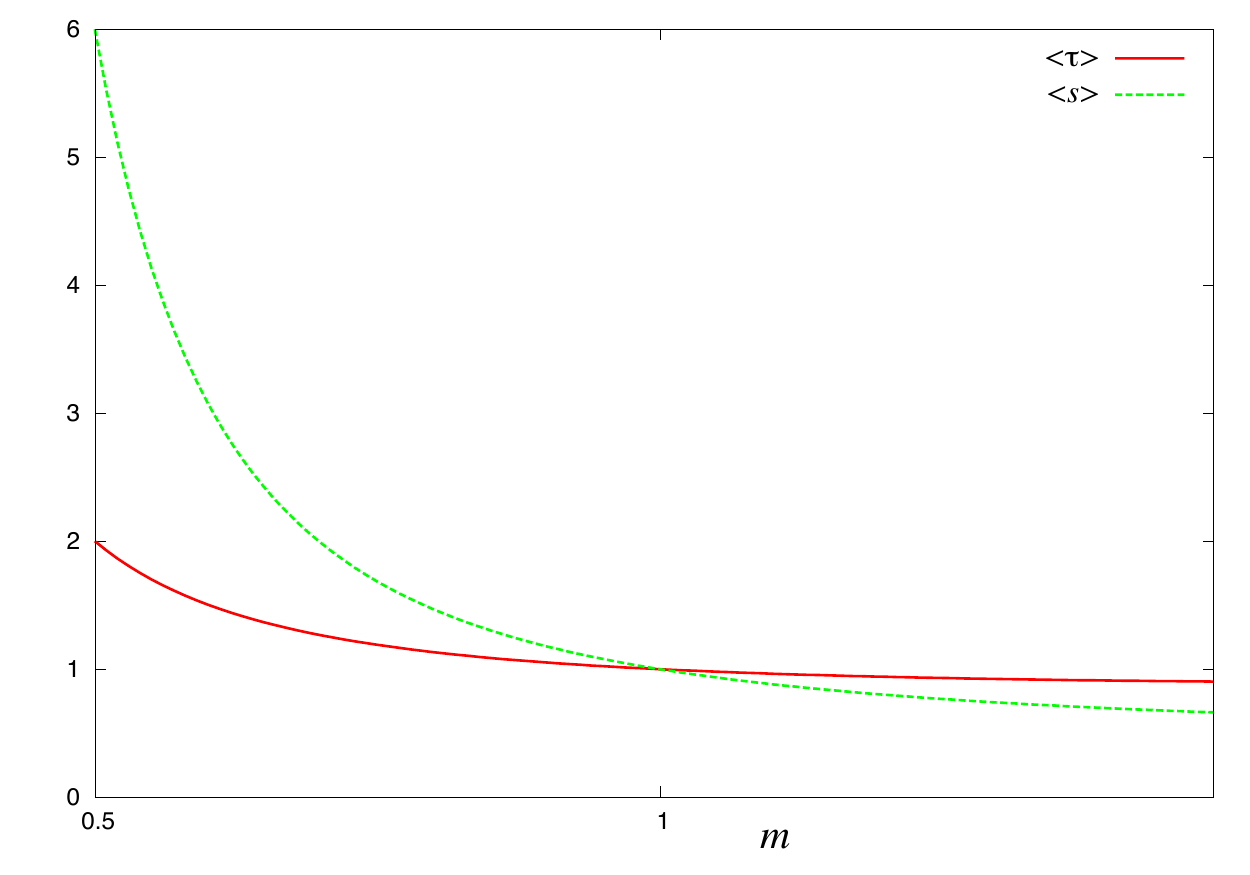}
\end{center}
\caption{\footnotesize 
Average duration $\langle \tau \rangle$ and 
average waiting time $\langle s \rangle$ as 
a function of $m$ 
for a Weibull duration distribution with $a=1$.  
The inspection paradox occurs for $m<m_{\rm c}=1$.}
\label{fig:inspect}
\end{figure}

Here we encounter the situation which is known as ``inspection paradox''.
For the Weibull distribution,
the paradox occurs for $m<m_{\rm c}=1$. 
Namely, for this regime, 
we have $\langle s \rangle > \langle \tau \rangle$ 
(see Fig. \ref{fig:inspect}).
In general, it means that the average of 
durations (first-passage times) is shorter than the average 
waiting time. 
This fact is quite counter-intuitive 
because the customer checks the rate 
at a time between arbitrary 
consecutive rate changes. 
This fact is intuitively understood as follows. 
When the parameter $m$ is smaller than $m_{\rm c}$, 
the bias of the duration is larger than that of 
the exponential distribution. 
As a result, the chance for customers to 
check the rate within large intervals between 
consecutive price changes 
is more frequent than the chance they check the rate 
within shorter intervals. 
Then, the average waiting time can become 
longer than the average duration. 
\section{Concluding remarks}
\label{sec:Summary}
As we showed in this paper, 
our queueing theoretical 
approach might be useful to build 
artificial markets such as the on-line trading 
service so as to have a suitable 
waiting time for the individual customers 
by controlling the width of the rate window.  
Moreover, the theoretical framework we provided here 
could predict the average waiting time including 
the deviation from empirical results. 

We hope that this review article might help 
researchers or financial engineers when they attempt to build 
a suitable on-line system for their customers. 

\end{document}